\documentstyle[12pt]{article}
\def\beq{\begin{equation}}
\def\eeq{\end{equation}}
\def\bea{\begin{eqnarray}}
\def\eea{\end{eqnarray}}
\def\bq{\begin{quote}}
\def\eq{\end{quote}}

\def\gappeq{\mathrel{\rlap {\raise.5ex\hbox{$>$}}
{\lower.5ex\hbox{$\sim$}}}}

\def\lappeq{\mathrel{\rlap{\raise.5ex\hbox{$<$}}
{\lower.5ex\hbox{$\sim$}}}}

\begin{document}
\topmargin -0.5cm
\oddsidemargin -0.8cm
\evensidemargin -0.8cm
\pagestyle{empty}
\begin{flushright}
UPR-578T\\ {CERN-TH.7051/93}
\end{flushright}
\vspace*{5mm}
\begin{center}
{\bf NON-PERTURBATIVE INTERACTIONS} \\
{\bf IN TWO-DIMENSIONAL} \\
{\bf (SUPER) STRING THEORY} \\

\vspace*{0.3cm}
{\bf Ram Brustein}$^{*),**)}$, {\bf Michael Faux}$^{*)}$ and  {\bf Burt A.
Ovrut}$^{*)}$\\ \vspace{0.3cm}
{\bf   Dept. of Physics, Univ. of Pennsylvania\\
Philadelphia, PA 19104, U.S.A.\\
and\\
Theoretical Physics Division, CERN \\
CH - 1211 Geneva 23\\}
\vspace*{0.3cm}
\vspace*{2cm}
{\bf ABSTRACT}
\end{center}
\vspace*{3mm}
\noindent
Non-perturbative interactions in the effective action of
two-dimensional bosonic string theory are described.
These interactions are due to ``stringy" instantons that are associated
with a space-varying coupling parameter. We present progress towards
identifying similar non-perturbative interactions in the effective
action of two-dimensional superstring theory. We discuss the possible relation
to higher dimensional string theory.

\vspace*{1.8cm}
\noindent
\rule[.1in]{16.5cm}{.002in}\\
\noindent
$^{*)}$ Work supported in part by DOE under Contract No.
DOE-AC02-76-ERO-3071.\\
$^{**)}$ Presented at International  Workshop on Supersymmetry
and Unification of Fundamental Interactions (SUSY 93), Boston, MA, 29
Mar - 1 Apr 1993.

 \vspace*{0.5cm}

\begin{flushleft} UPR-578T\\ CERN-TH.7051/93 \\
October1993
\end{flushleft}
\vfill\eject
\setcounter{page}{1}
\pagestyle{plain}
\font\tenrm=cmr10
\font\bigit=cmti10 scaled\magstep 2
\def\fl{\flushleft}
\def\L{{\cal L}}
\def\R{\cal{R}}
\def\a{\alpha}
\def\b{\beta}
\def\d{\delta}
\def\e{\epsilon}
\def\G{\Gamma}
\def\g{\gamma}
\def\l{\lambda}
\def\n{\eta}
\def\z{\zeta}
\def\k{\kappa}
\def\tPhi{\tilde{\Phi}}
\def\T{\theta}
\def\t{\theta}
\def\vphi{\varphi}
\def\w{\omega}
\def\ad{\dot{\alpha}}
\def\bQ{\bar{Q}}
\def\be{\bar{\epsilon}}
\def\bn{\bar{\eta}}
\def\bpsi{\bar{\psi}}
\def\bT{\bar{\T}}
\def\bD{\bar{D}}
\def\hf{\frac{1}{2}}
\def\der{\partial}
\def\bq{\begin{equation}}
\def\eq{\end{equation}}
\def\brr{\begin{eqnarray}}
\def\err{\end{eqnarray}}
\def\ba{\left(\begin{array}}
\def\ea{\end{array}\right)}
\def\pp{\hbox{\ooalign{$\displaystyle\int$\cr$-$}}}
\def\derbar{\stackrel{\leftrightarrow}{\partial}}
\def\dd{\stackrel{\leftrightarrow}{\partial}}
\def\ba{\left(\begin{array}}
\def\ea{\end{array}\right)}
\def\gbig {\hbox{\Large\it g}}
\input epsf
\vglue .8 cm
\vglue 0.6cm
{\bf\noindent 1. Introduction}
\vglue 0.4cm
To confront string theory with the observable world we have to  understand the
sources of non-perturbative interactions in the theory and know how to
calculate
them. Non-perturbative effects in string theory  are believed to determine a
number of important quantities,  for example,  the strength and type of
supersymmetry  breaking \cite{dixon}.  Matrix models  offer a unique
opportunity
to obtain some  insight into non-perturbative string theory.

Bosonic  d=1 matrix models \cite{done} have associated  with them  simple
string theories with a small number of degrees of freedom,  propagating in 1+1
space-time dimensions \cite{jevrev}.  An equivalent description is given in
terms of a bosonic collective field  theory\cite{das} in $1+1$ dimensions
involving  one massless field. Notable features of  collective field theory are
that the  kinetic energy  is not canonical and that the theory is not Poincare
invariant.  The coupling parameter of collective field theory varies in space
and becomes  strong in a certain region of space.  Yet another description is
given in terms of a canonical Poincare invariant effective field theory of two
fields with a field dependent coupling parameter\cite{ramburt}.  One of these
fields acquires a space dependent vacuum expectation value, which causes the
coupling parameter to become space dependent.

Natural candidates for describing the supersymmetric extension of the simple
string theories associated with bosonic matrix models are the theories
associated with supermatrix models \cite{marpar}. It turns out that the
construction of a 1+1 dimensional  supersymmetric effective  field theory
using
matrix models can be accomplished in several related ways. We illustrate the
various possibilities in figure 1. The boxes in this figure represent
intermediate steps in the construction of the supersymmetric effective theory.
The upper line represents the transformation from a bosonic matrix model
through
various associated bosonic theories.  The steps labeled 1-3  respectively
represent the extraction of the bosonic matrix eigenvalue theory through a
suitable restriction of the Hilbert space, the construction of the bosonic
collective field theory from the eigenvalue theory, and finally the
identification of a bosonic effective theory which is Poincare invariant and
which reduces to the bosonic collective field theory when a ``heavy" field is
frozen in its VEV.  The bottom line represents an analogous derivation starting
from a supersymmetric matrix model. Thus, the steps labeled $I-III$
respectively
represent the extraction of a supersymmetric eigenvalue theory by a suitable
restriction of the Hilbert space, the construction of the
associated collective field theory, and finally the identification of
an effective field theory which is both Poincare invariant and
supersymmetric.  The steps labeled $A-D$ represent direct supersymmetrizations
of the various bosonic theories.  Thus, a supersymmetric effective
theory can be obtained by supersymmetrizing the bosonic effective
theory.  Alternatively, any intermediate bosonic theory could
be supersymmetrized and then the remaining steps in the bottom
line could be followed until a supersymmetric effective theory
is obtained.
\ \vspace{1pc}

\epsfysize=113pt \centerline{\epsfbox{f0suin.epsf}}
\centerline{\tenrm Figure~1. Different ways to obtain a supersymmetric
            effective Lagrangian using matrix models}
 In this talk we emphasize the route, labeled in figure 1 by $A-I-II-III$,
from a bosonic matrix model to a supersymmetric effective theory.
However, at every step, it is possible to check that
our result is, in fact, the appropriate supersymmetrization of the
corresponding bosonic theory.  We have thus shown \cite{bfo} that the diagram
indicated in figure 1 commutes completely.  Previously, some of the steps
mentioned above have been analyzed with partial success
\cite{{dabh},{supjev},{sa},{jd},{jf}}.

The matrix models, or the equivalent field theories,
have the power to describe non-perturbative phenomena in the associated
$1+1$ (super) string theories.  Moreover, there are indications that  some
features of non-perturbative interactions in string theory  are common to all
string  theories, including more complicated superstring  theories in higher
dimensions  such as $d=4$. By studying the generic features of non-perturbative
behavior in $1+1$ dimensional (super) string theories,
one may therefore learn about more realistic four-dimensional (super) string
theories.

The first indication of   non-perturbative
behavior common to the simple theories in $1+1$ dimensions associated
with matrix models and more complicated string theories comes from large order
growth of perturbative amplitudes\cite{shenker}.
All matrix models, as well as critical bosonic string theory
in 26 dimensions, exhibit a strange phenomenon. The magnitude of $G$'th order
amplitudes in  perturbation theory  grow like $(2G)!$.
It turns out that, in much the same way  as $G!$ behavior corresponds to
$e^{-{1/g^2}}$ non-perturbative effects in quantum field theory,
in matrix models the large order  $(2G)!$ behavior
would correspond to non-perturbative effects of strength $e^{-{1/g}}$.
How do these peculiar effects arise?
In matrix models, there is a new type of instanton, involving
a single matrix eigenvalue, that is responsible for these effects.
{}From the point of view of the effective field theory, the existence of this
instanton is closely related to the space dependence of the coupling parameter.
Of particular importance is the fact that the  coupling parameter becomes
strong
in a certain region of space.

Another indication comes from studying  spherically symmetric
linear dilaton solutions in higher dimensional
superstring theory\cite{{ld},{caetal},{giddstro}}.
The main feature of these solutions is that  the coupling parameter of the
theory  varies in space and becomes strong in a certain region of space.
The effective  theory of  light fields in the background of these solutions
\cite{giddstro} is similar to the field theories associated with matrix models.
Since the existence of the above mentioned instantons is closely related to
having a coupling parameter that becomes strong in a certain region of space
and since this very feature characterizes the effective field theories for
light fields around some spherically symmetric superstring solutions,
we may expect that  non-perturbative interactions similar to those present in
the theories associated with matrix model are also present in higher
dimensional  (super) string  theory.

In this talk we review the results obtained for the bosonic theory. We present
 the supersymmetric extension of the bosonic theory and report about  progress
towards determining the nature of non-perturbative interactions in the
supersymmetric theory. Parts of this talk are based on the material contained
in  recent work \cite{ramburt,bfo}.
 More details as well as a comprehensive list of references can be found
in those papers.
\vglue 0.6cm
{\bf\noindent 2. Effective Action and Non-Perturbative Interactions
for Bosonic Strings}
\vglue 0.4cm
{\it\noindent 2.1. Collective Field Theory for Bosonic Matrix Models}
\vglue 0.4cm

Collective field theory for bosonic $d=1$ matrix models is written in terms
of the  density of eigenvalues,
$\vphi'(x,t)=\sum\limits_i\delta(x-\lambda_i(t))$ where $\lambda_i$
are the eigenvalues of the matrix and the prime denotes $\partial_x$.
 The size of the matrix, $N$, is very large.
The field $\vphi$ is called the  collective field.
The  Lagrangian  density  of collective field theory  is
\begin{equation}
{\cal L}=
{1\over 2}{\dot\vphi^2  \over\vphi'}
-{\pi^2\over 6} \vphi'^3+{1\over 2} x^2\vphi'
\label{collag}\end{equation}
Note that the field $\vphi$ does not have canonical kinetic energy
and that the theory is not Poincare invariant.
The classical equations of motion are not the usual field
equations\cite{ramburt}.  Instead  they are
\begin{equation}
\left[\partial_t{\dot\vphi\over\vphi'}
-\partial_x\!\left(\!{1\over 2} {\dot\vphi^2\over\vphi'^2}
 \!-\!{\pi^2\over 2}\vphi'^2
\!+\!{1\over 2} \!x^2)\right)\right]_{\!\!|x=\lambda_i(t)}
\hspace{-2pc}=0 \label{colleq}
\end{equation}
where  $i=1,...,N$.

These equations allow solutions in both high eigenvalue density regions  and
low
eigenvalue density regions. The static, high density, solution of these
equations  is very simple. It is $\vphi'_0=
{1\over\pi}\sqrt{x^2-{1\over g}}$,  where $|x|\ge\sqrt{{1\over g}}$ and the
parameter $1/g$ is an integration  constant.
There are also interesting time-dependent, Euclidean single eigenvalue
solutions
to Eq.(2). These are
\begin{equation}
\vphi^{(\pm)}_{inst}(x,\theta)=
\Theta\left(x\mp{1\over\sqrt{g}} \sin(\theta-\theta_0)\right)
 \label{pinst}\end{equation}
These  instantons describe tunneling of one eigenvalue
 across the barrier from
$x=\pm\sqrt{{1\over g}}$ at Euclidean time $\theta=\theta_0^{(\pm)}-\pi/2$ to
$ x= \mp\sqrt{{1\over g}}$ at $\theta=\theta_0^{(\pm)}+\pi/2$.
The action of these
instantons is $\pi/2 g$ in agreement with the large order growth of
perturbative amplitudes. One eigenvalue instantons have been  discussed
elsewhere in a different context. \cite{leemende}

\vglue 0.6cm
{\it\noindent 2.2. Effective Field Theory}
\vglue 0.4cm

The collective field theory Lagrangian density,  (\ref{collag}), has two
notable
deficiencies. First, the kinetic energy term is not in canonical form. This
means that we have not identified correctly the canonical field of the theory.
Second, and more important, the  coordinate $x$ appears  in the  potential
energy and therefore Poincare invariance is broken explicitly.  We
remove both deficiencies. The first, following  ref.\cite{das}, is
removed by  shifting and expanding around the classical solution
$\vphi'=\vphi'_0+{1\over \sqrt{\pi}}\partial_x\zeta$ and a coordinate
redefinition $x\rightarrow\tau={1\over\pi}\int\limits^x {dy\over\vphi'_0}$.
The
second, following ref.\cite{ramshanta}, is removed
 by enlarging the theory to include a new field, $D$. The non-trivial
vacuum expectation value of this  new field is responsible for the spontaneous
breaking of  Poincare invariance. The resulting action  for the fields
$D$ and $\zeta$ in the new coordinates is
\medbreak
\begin{eqnarray}
{\cal S}\!\!\!\! &=&\!\!\!\! \int dt d\tau \Biggl\{  {1\over 2}
{ \nabla \zeta\cdot\nabla \zeta \over 1+ 2\sqrt{\pi} g
   { {1\over\k} e^{D}\over\left(1- {1\over \k} e^{D}\right)^2}
\nabla\zeta\!\cdot\!\nabla D  }
-\!{\sqrt{\pi} g \over 4 }  {    {1\over\k} e^{D}
\over \left( 1- {1\over \k} e^{D} \right)^2 }
{(\nabla \zeta\cdot\nabla D)^3\over 1+ 2 \sqrt{\pi} g
{ {1\over \k} e^{D}\over \left(1-{1\over \k} e^{D} \right)^2}
 \nabla \zeta\cdot\nabla D}
\nonumber\\ &-& {\sqrt{\pi} g \over 12}
{ {1\over \k} e^{D} \over\left(1-{1\over \k} e^{D} \right)^2 }
(\nabla\zeta\cdot\nabla D)^3 -
{1\over  384 \pi} {\k^2\over g^2} e^{-2 D} \left[1-{1\over \k} e^{D} \right]^4
\left[ {(\nabla D)^2-4} \right]
\Biggr\} \label{efflag}
\end{eqnarray}
where $\k$ is an integration constant which relates the matrix model and
continuum space-time length scales. The action (\ref{efflag}) is not unique and
we discuss below some ambiguities in its form.  The kinetic energy for $\zeta$
is clearly canonical  and Poincare invariance is manifest.

The action is  invariant under
a somewhat  unexpected\cite{banks} shift symmetry
$\zeta\rightarrow\zeta+const$.
The field $\zeta$ has no potential, only derivative interactions.
 If one starts from collective field theory the reason for the appearance
of derivative interactions for the field $\zeta$ is obvious. This is a direct
consequence of  the conservation of the number of eigenvalues of the matrix or
of the equivalent condition,  $\int dx \vphi'(x)=N$, satisfied by the
collective
field. The way this condition is enforced on fluctuations
around a classical solution is by making their integral vanish. This means that
these fluctuations can be written as a derivative of some field $\zeta$.
This condition has to be enforced to all orders in perturbation theory.

The general solution of the equations of motion derived from (\ref{efflag})
is given by
\begin{equation}
<D>  =   \sinh\gamma_0 (t- t_0) + \cosh\gamma_0 (\tau -\tau_0),\ \
<\zeta>  =c \end{equation}
where
$c$, ${\gamma_0}$, $\tau_0$ and $t_0$ are arbitrary real parameters.
An interesting vacuum solution, which is a combination of two solutions
is $<\zeta>={1\over g}$ and  $<D>=-2\tau$ for $\tau\ge\ln\sqrt{\k}$,
henceforth called region $I$, and $<D>=+2\tau$
for $\tau\le-\ln\sqrt{\k}$, henceforth called
region $III$. When we  substitute the vacuum expectation value (VEV) of $D$
into
the Lagrangian (\ref{efflag}) thus freezing $D$ at its VEV and  shift the field
$\z$ around its VEV and transform the result back to $x$ space we recover the
collective field Lagrangian (1).

The coupling  parameter  of the effective theory,
\begin{equation}\hbox{\bigit g} (D)=4\sqrt{\pi}g
 { {1\over\k} e^{D}\over\left(1- {1\over {\k}} e^{D} \right)^2 }
\end{equation}
becomes space dependent once $D$ obtains the above expectation value
\bq \gbig_\pm(\tau)=
     4\sqrt{\pi}g\frac{\frac{1}{\k} e^{\mp 2\tau}}
     {(1-\frac{1}{\k}e^{\mp 2\tau})^2}
\label{gdef}\eq
We plot the effective coupling parameters in Figure 2.
 The effective coupling parameter in region $I$ is
${\hbox{\bigit g}}_-$
 and in region $III$ is ${\hbox{\bigit g}}_+$.
The spatial interval  $-\ln\sqrt{\k} \le \tau\le \ln\sqrt{{\k}}$
is called region $II$.
The two dashed ``walls" in Figure 2 mark the boundaries of the regions.
As can be seen from Figure 2,  in the regions far away from the two walls,
the coupling parameters are small and we expect the effective field theory
(\ref{efflag})
to provide a good description of physics. Near the ``walls" the coupling
parameter blows up.

\ \vspace{1pc}
\epsfysize=113pt \centerline{\epsfbox{f4suinn.epsf}}
\centerline{\baselineskip=12pt\tenrm Figure~2.
Effective coupling parameters in different regions of space.}
We use low density collective field theory as a
guide in  region $II$. Comparing the low density collective field theory
solution to the solution of the effective field theory we see that in region
$II$ their solutions should be the same.  In Euclidean space the solution is
therefore  the  instanton of Eq.(3),  expressed in the new coordinates
\begin{equation}\z_{inst}^{(\pm)}(\tau,\t) =
     \sqrt{\pi}\Theta\bigg{(}\frac{1}{\sqrt{g}}
[\sin(\frac{\pi}{\ln\k}\tau)\mp\sin(\t-\t_0^{(\pm)})]\bigg{)}\end{equation}
This is an instanton which describes  the tunneling of a single eigenvalue
across the barrier from $\tau=\pm\ln\sqrt{ \k}$ at  Euclidean time
$\theta=\theta_0-\pi/2$ to $\tau=\mp\ln\sqrt{\k}$ at $\theta=\theta_0+\pi/2$.

\vglue 0.6cm
{\it\noindent 2.3. Non-Perturbative Interactions }
\vglue 0.4cm

We  integrate over the instantons and represent
their effects as effective terms in the $D$, $\zeta$ theory.
The most general  action   induced by instantons is \\
$\Delta S=\int d\tau d\theta  \{\sum\limits_n C_n O_n(\tau,\theta)\}$,
where $O_n$ are local operators built from  $D$ and $\zeta$ and
their derivatives.
The coefficients $C_n$ can be computed by expanding the action around the
instanton background.
All the coefficients $C_n$ are proportional to the universal factor of the
exponent of the instanton action  and the remaining
factor depends on the  particular operator.  That is,
\begin{equation}C_n= \hbox{\it \~C}_n  e^{-{\pi\over 2 g} }\end{equation}
For example, the coefficient  of the unit operator is given by
$C_0= \hbox{\it \~C}_0 {\scriptstyle } e^{-{\pi\over 2 g} }$.
This result  was obtained in the background of a constant
field $<\zeta>={1\over g}$. Poincare invariance then dictates that at least for
slowly varying fields the effective operator depends on the full field $\zeta$
and not just its constant mode  ${1\over g}$.  Therefore the final result of
the  induced operator is
\begin{equation} \Delta{\cal L}_0=\hbox{\it \~C}_0(\zeta)
e^{-\frac{\pi}{2}\zeta}
\end{equation}
This operator breaks the  $\zeta$ shift symmetry. It induces a
non-perturbative
potential for the field $\zeta$. It is easy to understand why our instantons
break the shift symmetry. Recall that the shift symmetry  is related  to the
conservation of the number of eigenvalues of the matrix or to the equivalent
condition $\int dx \vphi'(x)=N$.  The eigenvalue associated with the instanton
``spends some of its time" inside the barrier and therefore the effective
number
of eigenvalues is reduced and the shift symmetry is broken.

\vglue 0.6cm
{\it\noindent 2.4. Bosonic String  Theory in Two Dimensions}
\vglue 0.4cm

The action (\ref{efflag}) is the effective space-time action that corresponds
to
string theories described by the following world sheet $\sigma$-model in
a special background.
\begin{equation}
I={1\over 4\pi}\int d^2z\sqrt{\hat g} \biggl\{ \hat g^{\a\b}G_{\mu\nu}
\partial_{\alpha}X^{\mu}\partial_{\beta}X^{\nu}+
\hat R D(X)+2 T(X)\biggr\}\label{smodel} \end{equation}
Here $\hat g_{\a\b}$ is the fixed world sheet metric  with Euclidean
signature and  $\hat R$ is the corresponding Ricci scalar.
The sigma model  field $X_\mu$  stands for  two  scalar
fields, $X_0 (z)$, and $X_1(z)$.
The   field $G_{\mu\nu}(X)$ is the target space metric,  assumed here  to  have
Euclidean signature, $D(X)$ is the dilaton,  and $T(X)$ is the tachyon.
The field $\zeta$ is related to the tachyon.

The relationship between the $\sigma$-model and the effective Lagrangian is
established in the following standard way . One first finds the  solutions of
the conformal invariance
conditions. In this case there exist  exact classical solutions. For example,
\begin{eqnarray}
G_{\mu\nu}^{(0)} & =&\ \eta_{\mu\nu}\nonumber\\
D^{(0)}\  & =&-2 X_1\nonumber\\ T^{(0)}\ &=& m e^{-2 X_1}\label{csol}
\end{eqnarray}
where $m$ is a constant. One now expands $T=T^{(0)}+\widetilde T$ and
 computes the on-shell scattering amplitudes for the massless
field $\widetilde T$ around this classical background\cite{frakut}. These
scattering amplitudes were shown to be the equal to those computed using the
matrix model and collective field theory in the region of momentum space in
which they can be compared. The equality holds provided the parameters $m$ and
$1/g$ are identified.  The collective field Lagrangian is therefore a
generating
functional for string amplitudes around the particular classical solution
(\ref{csol}). For a recent review and discussion of these issues, see
ref.\cite{jevrev}.  The effective Lagrangian, when expanded around its
classical solution is exactly equal to the collective field theory Lagrangian.
The effective field theory Lagrangian is therefore also a generating functional
for string amplitudes. The relationship between the fields $\z$ and $T$ can be
determined, asymptotically, in the region $X_1\rightarrow\infty$ to be
$\z\propto T e^{-D}$.

We are now in a position to state the ambiguities in the form of the action
(4). These are the usual ambiguities associated with the relationship between
on-shell scattering amplitudes and the Lagrangians generating them.
Terms that are identical on-shell cannot be distinguished. In particular,
this means that if take any of the $D$, $\z$ interaction terms  in (4) and
multiply them by  $-{\nabla D \nabla  D\over 4}$ or replace, for example, the
number 4 in the pure $D$ term by some power, $4 (-{\nabla D \nabla  D\over
4})^n$
 the effective Lagrangian would still be a good generating functional for
on-shell scattering amplitudes. Of course,  a symmetry (like supersymmetry) can
restrict the form of some of the terms even off-shell. In (4) we have chosen
the
terms with the least possible number of derivatives.

  \vglue 0.6cm
{\it\noindent 2.5. Possible Relation to String  Theory in
Higher Dimensions}
\vglue 0.4cm
The massless bosonic sector of  superstrings in ten dimensions
is described by the following action
\begin{equation}
\int d^{10}x \sqrt{-G} e^{-2 D}\left\{ R+4 (\nabla D)^2 -{1\over 3} H^2\right\}
\label{hetact}\end{equation}
where $H$ is the field strength of the  antisymetric tensor $B$, $D$ is the
dilaton and $R$ is the Ricci curvature scalar. There
are many interesting solutions to the equations of motion
derived from  the action (\ref{hetact}).
Some of them contain regions that are  approximately flat two-dimensional
space and have a  linear dilaton\cite{{caetal},{giddstro}}. Consider for
example
the five-dimensional extremal blackhole\cite{caetal}
\begin{eqnarray}
ds^2& =& -Q dt^2+{Q\over r^2} (dr^2 +r^2 d\Omega^2_3)   \nonumber\\
e^{2(D-D_0)}& =&  1+{Q\over r^2}  \nonumber\\
H&=& Q \epsilon_3
\end{eqnarray}
In the ``throat" region, ${Q\over r^2}>>1$ the solution is approximately
\begin{eqnarray}
ds^2& \sim& -Q dt^2 +Q d\tau^2 +Q d\Omega^2_3   \nonumber\\
D-D_0& \sim&  - \tau  \nonumber\\
H&=& Q \epsilon_3
\end{eqnarray}
where $\tau\sim \ln r$.
There are lots of variants
of this solution which exists in different dimensions including $d=4$
\cite{psgkkl}. It is important to note that the dilaton in the exact
solution is asymptotically constant. It varies linearly only in the ``throat"
region.

Consider the expansion
of the action (\ref{hetact}) around the classical solution in the ``throat"
 region\cite{giddstro}.  Because the geometry of this region is that of
$M_2\times S_3$, the light fields can be described by an effective
two-dimensional field theory in $(t,\tau)$ space. This theory is, of course,
not
Poincare invariant because the dilaton has a space dependent expectation value.
The coupling parameter of the theory varies exponentially in space for the same
reason.  The light fields of the $d=2$ theory can be found using standard
techniques (see for example ref.\cite{schwarz}). They include some modes of
the antisymmetric tensor which  can be described by  derivatively coupled
scalar
fields.

The similarity between our $d=2$ theory and the one associated with regions
of linear dilaton solutions of superstring is suggestive.
It may well be that this similarity is not accidental and that the effective
field theory derived from matrix models actually describes the linear dilaton
region of some exact solution to superstring equations of motion. If this is
indeed the case the results obtained using matrix models can be expected
to apply to  higher dimensional  superstring theories as well.

\vglue 0.6cm
{\bf\noindent 3. Effective Action  and Non-Perturbative Interactions
for SuperStrings}
\vglue 0.4cm
In this section we present the supersymmetric
extension of the bosonic effective theory and sources of
non-perturbative interactions  in this theory.

\vglue 0.6cm
{\it\noindent 3.1 Supersymmetric Matrix Model and Supersymmetric
Collective Field Theory}
\vglue 0.4cm

We extend the bosonic theory by letting the fundamental variable be a
time-dependent $N\times N$ matrix whose elements are $d=1, {\cal{N}}=2$ complex
superfields. The  Lagrangian for this  model is
\bq
 L=\int d\t_1d\t_2\Biggl\{\hf Tr D_1\Phi D_2\Phi+iW(\Phi)\Biggr\}
 \eq
where $\Phi=M+\bar\Psi\t+\bT\Psi+\t\bT F$ is a matrix superfield.
The matrices $M$ and $F$ and $\Psi$ are $N\times N$  matrices, $M$ and $F$
are Hermitian matrices and  $D=\der_\t+i\bT\der_t$.
The superpotential $W$ is an arbitrary
polynomial to be determined later.

One can  write an action for the singlet sector
of the theory in terms of the eigenvalues of $M$ and the diagonal elements
of $\Psi$ and $F$\cite{dabh}.  These can be written in terms of N superfields
$X_i=\lambda_i+\bar\chi_i\t+\bT\chi_i+\t\bT f_i$.
Equivalently, it is possible to write a collective field theory  to describe
the
singlet sector of the theory. In addition to the bosonic collective field
$ \vphi(x,t) = \sum_i\Theta(x-\l_i(t))$ there are two fermionic collective
fields $\psi(x,t) = -\sum_i\d(x-\l_i(t))\chi_{i}(t)$ and
     $\bar{\psi}(x,t) = -\sum_i\d(x-\l_i(t))\bar{\chi}_{i}(t)$.
The Lagrangian for the singlet sector can be written using the collective
fields as
\brr L &=& \int dx\Biggl\{
       \frac{\dot{\vphi}^2}{2\vphi'}
       -\hf\vphi'W'(x)^2
       +\frac{W''(x)}{\vphi'}\bar{\psi}\psi
       -\frac{1}{2\vphi'}(\bar{\psi}\dot{\bar{\psi}}
       +\dot{\bar{\psi}}\bar{\psi})
       +\frac{i}{2}\frac{\dot{\vphi}}{\vphi^{'2}}
       (\bar{\psi}\psi'-\bar{\psi}'\psi)\Biggr\}\nonumber \\
 &+&\frac{1}{3}\pp dxdydz\frac{\vphi'(x)\vphi'(y)\vphi'(z)}
       {(x-y)(x-z)} +\pp dxdy\frac{\vphi'(x)\vphi'(y)}{(x-y)}W'(x)\nonumber \\
 &+&\pp\frac{1}{(x-y)}\Biggl\{\bar{\psi}(x)\psi'(y)
       -\frac{\vphi''(y)}{\vphi'(x)}\bar{\psi}(x)\psi(x)\Biggr\}
 \label{lagcollpp} \err
We now have to define the high density continuum limit
of the theory.  This involves performing the principal part integrals,
$\pp dx \frac{1}{(x-y)}= \int dx \frac{1}{(x-y)}\pm i \pi \int dx \delta(x-y)$,
as well as  specifying the $N$ dependence of the  coupling parameters in the
matrix model superpotential $W$ (which is also referred to as
``taking the double scaling limit").
We perform the principle part integrals such that the resulting Lagrangian
is Hermitian in the new variables. We also must choose some ambiguous signs
along
the way. It turns out that  the most general superpotential for our purposes is
of the form
\bq W(x)=Nc_0+\sqrt{N}c_1x+\frac{1}{6}\frac{c_3}{\sqrt{N}}x^3
 \label{potform}\eq
Then $W'(x)^2 = Nc_1^2+c_1c_3x^2+\frac{1}{3}\frac{c_1c_4}{\sqrt{N}}x^3
     +\cdots$ and $W''(x) = \frac{c_3}{\sqrt{N}}x+\cdots$. We see that, as
$N\rightarrow\infty$,
$W''(x) \rightarrow 0$ and
 $W'(x)^2\rightarrow \Lambda-\w^2x^2$
where $\Lambda=\hf N c_1^2$ and $\w^2=-c_1c_3$ are positive
constants.  We can therefore neglect the third term in (\ref{lagcollpp}).
Furthermore,  it turns out that all subsequent results
are correct if we simply take $\Lambda=0$, take $\w=1$ and treat $\vphi$ as a
regular  unconstrained field. The Lagrangian then becomes
\brr L &=& \int dx\Biggl\{\frac{\dot{\vphi}^2}{2\vphi'}
     -\frac{\pi^2}{6}\vphi^{'3}
     +\hf x^2 \vphi' \nonumber \\
     & & -\frac{i}{2\vphi'}(\psi_1\dot{\psi}_1+\psi_2\dot{\psi}_2)
     -\frac{i\pi}{2}\psi_1\psi_1'
     +\frac{i\pi}{2}\psi_2\psi_2'
     +\frac{i}{2}\frac{\dot{\vphi}}{\vphi^{'2}}
     (\psi_1\psi_1'+\psi_2\psi_2')\Biggr\}
 \label{lagcollnoncan} \err
We can now change coordinates and shift around the classical solution, as in
the bosonic case, and obtain a canonical collective field theory
\brr L &=& \int d\tau\Biggl\{
       \hf(\dot{\z}^2-\z^{'2})
       -\frac{i}{\sqrt{2}}(\psi_+\dot{\psi}_+-\psi_+\psi_+')
       -\frac{i}{\sqrt{2}}(\psi_-\dot{\psi}_-+\psi_-\psi_-')
       \nonumber \\
       & & -\hf\frac{\gbig(\tau)\dot{\z}^2\z'}{1+\gbig(\tau)\z'}
       -\frac{1}{6}\gbig(\tau)\z^{'3}
       +\frac{i}{\sqrt{2}}\frac{\gbig(\tau)\z'}{1+\gbig(\tau)\z'}
       (\psi_+\dot{\psi}_++\psi_-\dot{\psi}_-) \nonumber \\
       & & +\frac{i}{\sqrt{2}}\frac{\gbig(\tau)\dot{\z}}
       {(1+\gbig(\tau)\z')^2}(\psi_+\psi_+'+\psi_-\psi_-')\Biggr\}
       +\frac{1}{3}\int d\tau\frac{1}{\gbig(\tau)^2}
 \label{lagcollfinal} \err
where now the prime means $\der/\der\tau$ and
$\psi_{+,-}=\frac{2^{1/4}}{\sqrt{\pi}}\psi_{1,2}$
 The bosonic terms of this Lagrangian are
identical to the canonical bosonic collective field theory.
The coupling parameter of the theory is therefore given in Eq.(\ref{gdef})
and takes the familiar form  depicted in figure 2.

\vglue 0.6cm
{\it\noindent 3.2 Effective Field Theory, Supersymmetry
 and Non-perturbative  Interactions}
\vglue 0.4cm
The Lagrangian (\ref{lagcollfinal}) has obvious deficiencies.
It is not  Poincare invariance  and it is not supersymmetric.  We now pose the
question: Is it possible to interpret the collective field theory Lagrangian as
originating from a Poincare invariant, supersymmetric field theory?
We proceed to answer this question in the affirmative.
The Lagrangian (\ref{lagcollfinal}) has kinetic energy terms for both
the bosonic field, $\z$ and for the fermionic fields,
$\psi_\pm$, which are canonically normalized for a flat
two-dimensional space-time.  The interaction terms, however,
involve an explicit spatially-dependent coupling, $\gbig(\tau)$,
which violates Poincare invariance.  As in the bosonic theory, we interpret
$\gbig(\tau)$ to be the VEV of a function of an
additional ``heavy" field, related to $D$,  which we denote by $\a$.
Furthermore,
we infer the existence of an effective theory involving $\a$, as well as $\z$,
$\psi_+$ and $\psi_-$, which reproduces (\ref{lagcollfinal}) when $\a$ is
replaced by its $\tau$-dependent VEV. Additionally, we postulate that the
effective theory possesses a two-dimensional supersymmetry.  It follows that,
in
addition to $\a$, we must introduce its fermionic superpartners, $\chi_+$ and
$\chi_-$ which, of course, have vanishing VEV's. The field $\a$ and its
superpartners $\chi_+$ and $\chi_-$ are assumed to be heavy. We do not consider
their fluctuations but rather treat them as frozen in their VEV's.
It turns out, after extensive calculations, that, of all the possible $d=2$
(p,q) supersymmetries,  it is only possible to construct a (1,1) supersymmetric
theory.

We then have at our disposal  two superfields, $\Phi_1$ and $\Phi_2$,
\brr \Phi_1 &=& \z+i\t^+\psi_++i\t^-\psi_-+i\t^+\t^-Z  \nonumber \\
     \Phi_2 &=& \a+i\t^+\chi_++i\t^-\chi_-+i\t^+\t^-A.
 \err
to construct a (1,1) supersymmetric Lagrangian that  reproduces the high
density
collective field Lagrangian, (\ref{lagcollfinal}), when the heavy fields are
replaced by their VEV's.
Using these two superfields and the differential operators,
$D_+, D_-, \der_+$, and $\der_-$, we  have built the effective
theory piecemeal, order by order in the coupling, $\gbig(\tau)$.
We present here only the final result, the all orders (1,1) supersymmetric
effective Lagrangian,
\brr \L^{(eff)} &=&
     \int d\t_+d\t_-\Biggl\{
     D_+\Phi_1D_-\Phi_1
     +F_1(\Phi_2)D_+\Phi_2D_-\Phi_2
    -F_2(\Phi_2)\der_-D_+\Phi_2\der_+D_-\Phi_2 \nonumber \\
     & &
     -\frac{f(\Phi_2)}{\Phi_2^3}
     \frac{\der_{(+}\Phi_1\der_{-)}\Phi_2
           \der_{[+}\Phi_1\der_{-]}\Phi_2}
          {1+\frac{f(\Phi_2)}{\Phi_2}
           \der_{(+}\Phi_1\der_{-)}\Phi_2}
           D_{(+}\Phi_1D_{-)}\Phi_2
     +\frac{1}{3}\frac{f(\Phi_2)}{\Phi_2^5}
     (\der_{[+}\Phi_1\der_{-]}\Phi_2)^3
     D_+\Phi_2D_-\Phi_2 \nonumber \\
     & & \hspace{.6in}
     -\frac{f(\Phi_2)}{\Phi_2^5}
     \frac{(\der_{[+}\Phi_1\der_{-]}\Phi_2)^2
     \der_{(+}\Phi_1\der_{-)}\Phi_2}
     {1+\frac{f(\Phi_2)}{\Phi_2}
     \der_{(+}\Phi_1\der_{-)}\Phi_2}
     D_+\Phi_2D_-\Phi_2 \Biggr\}
 \label{effsusylag} \err
where
\brr F_1(\Phi_2) &=& -\frac{1}{48\pi\k g^2}
     (\frac{11}{5}\frac{\k^3}{\Phi_2^6}
     -\frac{28}{3}\frac{\k^2}{\Phi_2^4}
     +18\frac{\k}{\Phi_2^2}
     -4
     +\frac{5}{3}\frac{\Phi_2^2}{\k}) \nonumber \\
     F_2(\Phi_2) &=&  -\frac{1}{48\pi\k g^2}
     (-\frac{2}{5}\frac{\k^3}{\Phi_2^6}
     +\frac{8}{3}\frac{\k^2}{\Phi_2^4}
     -12\frac{\k}{\Phi_2^2}
     -8
     +\frac{2}{3}\frac{\Phi_2^2}{\k}) \nonumber \\
     f(\Phi_2) &=& 4\sqrt{\pi}g\frac{\frac{1}{\k}\Phi_2^2}
         {(1-\frac{1}{\k}\Phi_2^2)^2}
 \err
The equations of motion derived from (\ref{effsusylag}) are satisfied
by the solution
\brr <\a> &=& \exp\Biggl\{
     [\sinh\gamma_0|t-t_0|-\cosh\gamma_0|\tau-\tau_0|]\Biggr\}
     \nonumber \\
     <\z> &=& constant \nonumber \\
     <\chi_\pm> &=& \n_0^\pm<\a>
 \label{solongen} \err
with all  other fields vanishing.
This solution is labeled by the
translational zero modes $t_0$ and $\tau_0$, a Lorentz zero mode
$\gamma_0$ and by supersymmetric zero modes $\n_0^\pm$.  In a preferred
frame of reference, $\tau_0=\gamma_0=\n_0^+=\n_0^-=0$, and the solution
(\ref{solongen}) becomes equivalent to
\brr <\a> &=& \exp{(-|\tau|)} \nonumber \\
     <\z> &=& constant
 \label{solon} \err
with all other fields vanishing.
 If we substitute the solution (\ref{solon}) into (\ref{effsusylag}),
thus freezing the ``heavy" fields $\a, \chi_+$ and $\chi_-$ at their
VEV's, and shift the light fields $\z, \psi_-$ and $\psi_+$
around their VEV's, we recover the collective field Lagrangian
(\ref{lagcollfinal})
derived from the $d=1, {\cal N}=2$ supersymmetric matrix
model. If we compare now the supersymmetric Lagrangian to the bosonic
Lagrangian we see that $\a=e^D/2$.

The fact that the Lagrangian (\ref{effsusylag}) exists is quite remarkable.
Lets
pause for a moment to reflect on this. Our starting point was a one-dimensional
theory with $N$ (recall that $N\rightarrow\infty$) multiplets of ${\cal N}=2$
supersymmetry. We then, through a rather complicated procedure, managed to
combine the infinite number of one-dimensional multiplets into two multiplet of
a two-dimensional (1,1) supersymmetry. It is obvious that the two-dimensional
supersymmetry is closely related to the one-dimensional supersymmetry but the
exact relation is not clear to us at this moment.

The coupling parameter of the supersymmetric theory is identical
to the coupling parameter of the bosonic theory. It blows up
at the boundaries of a region centered at $\tau=0$ with width
$\ln\k$. Outside of this region, the high density collective field theory
(\ref{lagcollfinal}) is valid.
Within the region $|\tau|<\ln\k/2$ or, equivalently, in the
region $|x|<1/(\sqrt{g})$ however, as in the bosonic theory,  the collective
field theory must describe a finite number of eigenvalues. Then the appropriate
form for the collective field theory is given in (\ref{lagcollpp}).  This
Lagrangian is completely equivalent to the original eigenvalue Lagrangian,
which
is, in most cases, easier to use.  We now  present the instanton
configurations corresponding to solutions to the Euclidean equations of motion
of
this low density eigenvalue theory. They are
\brr \z^{(\pm)}(\tau,\t) &=&
     \sqrt{\pi}\Theta\bigg{(}\frac{1}{\sqrt{g}}
     [\sin(\frac{\pi}{\ln\k}\tau)\mp\sin(\t-\t_0^{(\pm)})]\bigg{)}
     \nonumber \\
     \psi_1^{(\pm)}(\tau,\t) &=&
     \d\bigg{(}\frac{1}{\sqrt{g}}
     [\sin(\frac{\pi}{\ln\k}\tau)\mp\sin(\t-\t_0^{(\pm)})]\bigg{)}
     \n_{10}^{(\pm)} \nonumber \\
     \psi_2^{(\pm)}(\tau,\t) &=&
     \d\bigg{(}\frac{1}{\sqrt{g}}
     [\sin(\frac{\pi}{\ln\k}\tau)\mp\sin(\t-\t_0^{(\pm)})]\bigg{)}
     \n_{20}^{(\pm)}
 \err
Note that  because of supersymmetry there are fermionic zero modes associated
with the instantons. These fermionic zero modes are essential for understanding
how one eigenvalue instantons affect supersymmetry.

It is clear that the instantons induce non-perturbative operators
in the supersymmetric effective theory. It is also expected on
the basis of general arguments that these would break the one-dimensional
supersymmetry. The most interesting question is, however, what are the effects
of the instantons  on the  two-dimensional supersymmetry?
We are now vigorously investigating the structure of the induced operators
and the fate of the two-dimensional supersymmetry.
 \vglue 0.6cm
{\bf \noindent 5. Acknowledgments \hfil}
\vglue 0.4cm
We would like to thank E. Witten for useful observations.
\medbreak



\begin{thebibliography}{9}
\bibitem{dixon}
L. Dixon, ``Supersymmetry breaking in string theory",
presented at the 15th APS Div. of
Particles and Fields Conf., Houston, TX,
January 3-6, 1990;\\
J. Louis, ``Status of supersymmetry breaking in string theory", presented at
Particle and Fields '91 conf. Aug 18-22, 1991, Vancouver, BC, Canada.

\bibitem{done}
 D. J. Gross and N. Miljkovic,
  {\it Phys.Lett.} {\bf B238} (1990) 217;\\
 P. Ginsparg and J. Zinn-Justin,
{\it Phys. Lett.} {\bf B240} (1990) 333 ;\\
 E. Brezin, V. Kazakov, Al. Zamolodchikov,
{\it Nucl. Phys.} {\bf B338} (1990) 673.

\bibitem{jevrev}
A. Jevicki, preprint, HET-918/TA-502 (1993).

\bibitem{das}
S. R. Das and A. Jevicki,
{\it Mod. Phys. Lett.}{\bf A5} (1990) 1639.

\bibitem{ramburt}
R. Brustein and B. Ovrut, {\it Phys. Lett.} {\bf  B309} (1993)
45;\\ R. Brustein and B. Ovrut, preprint, UPR-523T (1992).

\bibitem{marpar}
E. Marinari and G. Parisi, {\it Phys. Lett.} {\bf B247} (1990) 537.

\bibitem{bfo}
R. Brustein, M. Faux and B. A. Ovrut, preprint, CERN-TH.7013/93 (1993);\\
R. Brustein, M. Faux and B. A. Ovrut, preprint, CERN-TH.7017/93 (1993).

\bibitem{dabh}
A. Dabholkar, {\it Nucl. Phys.} {\bf B368} (1992) 293.

\bibitem{supjev}
A. Jevicki and J. P. Rodrigues,
{\it Phys. Lett.} {\bf  B268} (1991) 53.

\bibitem{sa}
J. P. Rodrigues and J. van Tonder, preprint, CNLS-92-02 (1992).

\bibitem{jd}
J. Cohn and H. Dykstra, {\it Mod. Phys. Lett.} {\bf A7} (1992) 1163.

\bibitem{jf}
J. Feinberg,  preprint, TECHNION-PH-92-35 (1992).

\bibitem{shenker} S. H. Shenker,
``The strength of non-perturbative effects in string theory",
presented at the Cargese Workshop on Random Surfaces, Quantum Gravity
and Strings, Cargese, France, May 28 - Jun 1, 1990.



\bibitem{ld}
R. Myers,  {\it Phys. Lett.} {\bf B199} (1987) 371;\\
I. Antoniadis, C. Bachas, J. Ellis, D. Nanopoulos,
{\it Phys.Lett.}{\bf B211} (1988) 393;\\
S. P. De Alwis,  J. Polchinski,  R. Schimmrigk,
{\it Phys. Lett.} {\bf  B218} (1988) 449.

\bibitem{caetal}
C. Callan, J. Harvey, A. Strominger,
{\it Nucl. Phys.} {\bf B359} (1991) 611.

\bibitem{giddstro}
S. B. Giddins and A. Strominger {\it Phys. Rev.} {\bf D46} (1992) 627.

\bibitem{leemende}
J. Lee and P. Mende, {\it Phys. Lett.} {\bf B312} (1993) 433.

\bibitem{ramshanta}
R. Brustein and S. P. De Alwis,
{\it Phys.Lett.} {\bf B272} (1991) 285.

\bibitem{banks}
T. Banks, {\it Nucl. Phys.} {\bf B361} (1991) 166.

\bibitem{frakut}
P. Di Francesco and D. Kutasov,{\it Nucl. Phys.}
{\bf B375} (1992) 119.

\bibitem{psgkkl}
S.B. Giddings, J. Polchinski and A. Strominger, preprint,  NSF-ITP-93-62
(1993);\\ E. Kiritsis, C. Kounnas and D. Lust, preprint, CERN-TH.6975/93
(1993).

\bibitem{schwarz}
J. H.Schwarz,  ``Dilaton-Axion symmetry", Presented at International Workshop
on String Theory, Quantum Gravity and the Unification of Fundamental
Interactions, Rome, Italy, 21-26 Sep 1992.

\end{thebibliography}
\end{document}